\documentclass[preprint,review,12pt]{elsarticle}

\usepackage{arydshln}
\usepackage{verbatimbox}
\usepackage{graphicx}
\usepackage{url}
\usepackage{hyperref}
\usepackage{amsmath,amsfonts}
\usepackage[linesnumbered,lined,boxed,commentsnumbered,ruled]{algorithm2e}
\usepackage{textcomp}
\usepackage{multirow}
\usepackage{multicol}
\usepackage{color}
\usepackage{natbib}
\usepackage{caption}
\usepackage{setspace}
\usepackage{comment}
\usepackage{threeparttable}
\usepackage{booktabs}
\usepackage{tikz}
\usetikzlibrary{arrows,positioning}
\usepackage{lipsum, adjustbox}

\usepackage{siunitx}
\usepackage{subcaption}

\newcommand{\tcolR}{\textcolor{black}}
\newcommand{\tcolB}{\textcolor{black}}

\journal{CASCON'21}

\begin{document}

\title{Text Classification for Predicting Multi-level Product Categories}

\author[1]{Hadi Jahanshahi}
\ead{hadi.jahanshahi@ryerson.ca}
\author[1]{Ozan Ozyegen}
\author[1]{Mucahit Cevik}
\author[2]{Beste Bulut}
\author[2]{Deniz Yigit}
\author[2]{Fahrettin F. Gonen}
\author[1]{Ay\c{s}e Ba\c{s}ar}

\address[1]{Data Science Lab at Ryerson University, Toronto, Canada}

\address[2]{Getir Perakende Lojistik A.S., Istanbul, Turkey}

\begin{abstract}
In an online shopping platform, a detailed classification of the products facilitates user navigation.
It also helps online retailers keep track of the price fluctuations in a certain industry or special discounts on a specific product category. 
Moreover, an automated classification system may help to pinpoint incorrect or subjective categories suggested by an operator.  
In this study, we focus on product title classification of the grocery products. 
We perform a comprehensive comparison of six different text classification models to establish a strong baseline for this task, which involves testing both traditional and recent machine learning methods.
In our experiments, we investigate the generalizability of the trained models to the products of other online retailers, the dynamic masking of infeasible subcategories for pretrained language models, and the benefits of incorporating product titles in multiple languages.
Our numerical results indicate that dynamic masking of subcategories is effective in improving prediction accuracy.
In addition, we observe that using bilingual product titles is generally beneficial, and neural network-based models perform significantly better than SVM and XGBoost models.
Lastly, we investigate the reasons for the misclassified products and propose future research directions to further enhance the prediction models.
\end{abstract}


\maketitle
\section{Introduction}

E-commerce platforms have been increasingly popular over the years.
The interest in e-commerce has only increased with the COVID-19 pandemic, which resulted in the proliferation of e-commerce companies \citep{gao2020buy}.
This in turn increased the competition in the e-commerce field, and lead to significant investments by the companies to enhance their platforms.

To facilitate user navigation, e-commerce platforms list their items within appropriate categories. 
From the sellers' perspective, proposing the appropriate category for a product given its description can be cumbersome and time-consuming. 
The complexity of the task increases further by the introduction of multi-level categorization. 
For instance, a milk product can be categorized under the dairy category and the milk subcategory. 
As the number of products sold in an e-commerce platform increases, it becomes more difficult for the online platforms to keep track of hundreds or thousands of categories. 
The product category classification models aim to automate finding the right category for a given product. 
In most cases, the only available information is the product title and description.

To automatically categorize the products, online retailers can use product category classification models instead of manually scanning all categories to find the suitable one for each product.
At first glance, the product title classification problem can be considered as a variant of the widely studied text classification problems. 
While there are certain similarities between these two problems, in the product category classification problem, input titles might be significantly different in terms of length of each instance, the distribution of lengths, and the grammatical structure of the input~\cite{yu2012product}. 
Thus, various strategies have been proposed to extract the most information from short texts. 
These strategies include using context-relevant concept word embeddings~\cite{xu2020incorporating}, using both word-level and character-level features to capture fine-grained subword information~\cite{wang2017combining}, using word-cluster embeddings~\cite{shen2018improving} and data augmentation~\cite{rosario2017data}.

\tcolR{Some examples of product titles and corresponding category and subcategory labels are shown in Table~\ref{tab:examples}. Such a product categorization has three significant benefits to an online retailer.
First, it can assist buyers in navigation through online platforms. 
A high-quality categorization of the products results in a more efficient and satisfactory user experience.
Second, it allows online retailers to control sales and marketing operations in a more organized manner. 
They can easily add new products to their system and track aggregated information about various product categories in real-time.
Finally, online retailers can classify and trace available products of other online retailers. 
Using the predictions of product title classification models on other online retailers' datasets, companies can track aggregated information about the availability of various product categories.} 



\setlength{\tabcolsep}{9pt}
\renewcommand{\arraystretch}{1.13}
\begin{table*}[!ht]
    \caption{Examples of products titles, category and subcategory labels from the training dataset (brand names are underlined)} \label{tab:examples}
    \resizebox{\linewidth}{!}{
        \begin{tabular}{lllll}
        \toprule
        \textbf{Product Title} & \textbf{Category} & \textbf{Subcategory} \\
        \midrule
        \textbf{\underline{\textit{Dardanel}} canned tuna with beans 185 gram} & Food & Canned Food \& Pickle  \\
        \textbf{\underline{\textit{Bounty}} 57 gram} & Snack & Chocolate \\
        \textbf{\underline{\textit{Tadim}} white chickpeas 200 gram} & Snack & Dried Nuts \& Fruits \\
        \textbf{\underline{\textit{Dr Oetker}} whipped cream 75 gram} & Basic Food & Cake Ingredients \\
       \textbf{\underline{\textit{Evim}}} & Home \& Living & Newspaper \& Magazine \\
        \bottomrule
        \end{tabular}
    }
\end{table*}

\paragraph{Research Goals} We consider a specific application of product title classification over grocery products. We aim to study \textit{grocery product title classification} tasks by designing a comprehensive experimental study. We test various machine learning models and methods, including both traditional and recent Natural Language Processing (NLP) methods. \tcolB{This research facilitates the categorization of new products, tracking other retailers' products in an aggregated form, and determining incorrectly classified products in the system.}

\paragraph{Contributions}
We summarize the contributions of our study as follows.
\begin{itemize}
    \item To the best of our knowledge, this is the first study that focuses on grocery product title classification.  
    
    \item We perform a comprehensive comparison of six different text classification models on the grocery product title classification task. 
    Our experiments establish a strong baseline for this task by testing both traditional and recent NLP methods.
    
    \item We investigate several strategies such as leveraging product titles in multiple languages and dynamically masking the infeasible subcategories for pretrained language models to obtain better predictive performance for the product title classification task.
    
    \item We measure the generalizability of product title classification models by evaluating the trained model on six different datasets obtained from different online retailers.
    
    \item We identify the challenges of grocery product title classification through a detailed analysis of the model predictions.
\end{itemize}

\paragraph{Structure of the Paper}
The rest of the paper is organized as follows. 
Section~\ref{sec:lit_rev} provides a brief discussion on the relevant literature on hierarchical product category classification, which is followed up by the description of employed methodology and datasets in Section~\ref{sec:methodology}.
Section~\ref{sec:experiments} explores the results of within- and cross multilevel product category classification as well as insights into the products that the models fail to predict. 
Finally, Section~\ref{sec:threats} describes the limitations and threats to validity, and Section~\ref{sec:conclusion} provides concluding remarks along with future research directions.

\section{Literature Review}\label{sec:lit_rev}
Hierarchical product category classification is a challenging task.
It requires product instances to be carefully assigned to multiple levels of categories. 
Over the past years, the interest in this problem has increased with the rise of online shopping and the availability of large datasets.

\citet{yu2012product} provide one of the first studies in this area. 
They conduct an extensive numerical study to illustrate how linear SVMs can be used for large-scale multi-class title classification and identify the differences between product title classification and text classification. 
They use a dataset from a large internet company, which contains 29 classes, and propose a multi-class SVM model for the classification task.
They also compare the effectiveness of different feature representations. 
Their experiments show that stemming and stop word removal are harmful and bigrams are effective.

There have been significant improvements in NLP models over the past decade. 
For word representations, methods such as Glove~\cite{pennington2014glove} and Word2Vec~\cite{mikolov2013efficient} became increasingly popular. 
More recently, advanced NLP models such as BERT~\cite{devlin2018bert}, ROBERTA~\cite{liu2019roberta} and XLM~\cite{ma2020xlm} have been shown to achieve state-of-art for many language tasks. 
These models, also known as pre-trained language models (PTMs), are used slightly differently compared to the previous machine learning models that are considered for the NLP tasks. 
They are first trained on large-scale unlabeled corpora to leverage a good understanding of natural language.
Then, depending on the task, a few layers are attached to the end of the ``pre-trained'' base model. 
Afterwards, the full network is fine-tuned end-to-end on a smaller task-specific corpus.
There are additional advantages of using PTMs over the traditional methods. 
The same base model can be used for many NLP tasks with computationally inexpensive task-specific fine-tuning. Furthermore, for most cases, a small hyperparameter tuning setup that includes a range of batch sizes, learning rates, and the number of epochs is recommended for fine-tuning these models~\cite{devlin2018bert}. 

The adoption of pre-trained language models can also be seen in the most recent work on the product category classification domain~\cite{zhang2020mwpd2020,tagliabue2021sigir}. 
Most of the recent literature in product category classification problems can be found in the ``Semantic Web Challenge'' competition and case studies published by the competing teams~\cite{zhang2020mwpd2020}. 
The second part of the challenge focuses on multi-level product category classification. 
The considered dataset in the competition contains more than 15,000 product instances randomly sampled from 702 vendors' websites. 
The products are labeled in GPC hierarchy~\footnote{https://www.gs1.org/standards/gpc}. 
As baseline models, teams tested the same configuration proposed by~\citep{tagliabue2021sigir} which uses the FastText algorithm. 
For evaluating the results, standard metrics such as Precision, Recall, and F1 score are used, and 
to measure the overall performance, Weighted-Average macro-F1 (WAF1) scores are reported by each of the participating teams. 
All the top submissions ended up using variants of the  BERT~\cite{devlin2018bert} architecture. 
For instance, \citet{zahera2020probert} (the Team DICE) proposed a multi-label BERT architecture called ProBERT for the multi-label product category classification. 
ProBERT contains fully-connected neural layers with Sigmoid activations for each classification task.
The winner of the competition, \citet{tencent2020mwpd} (Team Rhinobird) proposed a slightly more complex method, which uses BERT as the base model. 
To obtain a semantically rich representation, they use hidden states from the last hidden layers of BERT, resulting in 17 different BERT models. 
These models are then combined using a two-level ensemble strategy. 
In the first level, they apply five-fold cross-validation by splitting the training data into training and validation sets. 
Then, they train the same model five times, each time using a different fold as the validation dataset and the remaining folds as the training dataset. 
Afterwards, they average the probability outputs of these five models with the same model architecture but trained on a different dataset. 
In the second level, an ensemble of 17 different BERT models is created where each model votes for the prediction, and the most voted class is selected as the final prediction. 
Moreover, they propose a Dynamic Masked Softmax function that explicitly considers the dependencies among different category levels~\cite{zhang2020mwpd2020}. 
The dynamic masking of the subcategories based on the predicted category reduces the complexity of the optimization problem by filtering out the child categories unrelated to the predicted parent category.

To the best of our knowledge, our work constitutes the first study on multi-level classification for predicting grocery products categories. 
We create an extensive list of text classification models to address this problem. 
Unlike previous works, we leverage bi-lingual models to improve prediction performance based on Turkish and English product titles. 
Finally, we discuss the challenges of the grocery product category classification task through our detailed numerical study with datasets from different online retailers.

\section{Methodology}\label{sec:methodology}
In this section, we briefly discuss the datasets and the methods used for our multilevel product category classification task. 
Moreover, we provide more details on our experimental setup, including the evaluation metrics and parameter settings.

\subsection{Datasets}
Our datasets are obtained from online grocery markets in Turkey.
Specifically, we mined product information from seven online grocery websites in this domain. 
We use one platform as the training set and others as the test sets. 
Table~\ref{tab:datasets} shows the number of unique products that are mined from these websites. 
The mining phase is done at different times of the year to ensure all products sold by companies are included. As there are inconsistencies in the category and subcategory naming for different websites, we include only those products from testing sets whose categories and subcategories are available in the training set.
The information related to the number of products, categories, and subcategories before and after the cleaning phase is shown in Table~\ref{tab:datasets}. 
We select a medium-size dataset to test the performance of the algorithm. 
This process can be replicated using any other test set as the training set. 
However, we do not aim to incorporate the most comprehensive ones, e.g., Test Set-5, since the model may indicate a performance level that is not generalizable to other datasets. 
Another factor in choosing this specific dataset is its multilingual platform. 
As we aim to examine the model performance when using both Turkish and English titles, we choose to select a platform that provides this particular feature. 

\setlength{\tabcolsep}{4pt} 
\renewcommand{\arraystretch}{1.2} 
\begin{table*}[!ht]
    \caption{Datasets' descriptions} \label{tab:datasets}
    \resizebox{\linewidth}{!}{
    \begin{tabular}{l|rrr|rrr}
    \toprule
    \multirow{2}{*}{\textbf{Datasets}} & \multicolumn{3}{c}{\textbf{Before cleaning}} & \multicolumn{3}{|c}{\textbf{After cleaning}} \\
     & \multicolumn{1}{c}{\textbf{\# of products}} & \multicolumn{1}{c}{\textbf{\# of categories}} & \multicolumn{1}{c}{\textbf{\# of subcategories}} & \multicolumn{1}{|c}{\textbf{\# of products}} & \multicolumn{1}{c}{\textbf{\# of categories}} & \multicolumn{1}{c}{\textbf{\# of subcategories}} \\
     \midrule
    \textbf{Training set} 
    & 3,119 & 18 & 113 & - & - & * \\
    \midrule
    \textbf{Test Set - 1} 
    & 2,039 & 15 & 81 & 284 & 3 & 7 \\
    \textbf{Test Set - 2} 
    & 3,094 & 27 & 142 & 1,891 & 16 & 90 \\
    \textbf{Test Set - 3} 
    & 659 & 17 & 84 & 656 & 15 & 81 \\
    \textbf{Test Set - 4} 
    & 2,435 & 22 & 120 & 1,235 & 13 & 50 \\
    \textbf{Test Set - 5} 
    & 6,981 & 26 & 366 & 378 & 2 & 7 \\
    \textbf{Test Set - 6} 
    & 3,892 & 20 & 114 & 995 & 8 & 27 \\
    \bottomrule
    \end{tabular}}
\end{table*}

\begin{figure}[!ht]
  \centering
    \includegraphics[width=\linewidth]{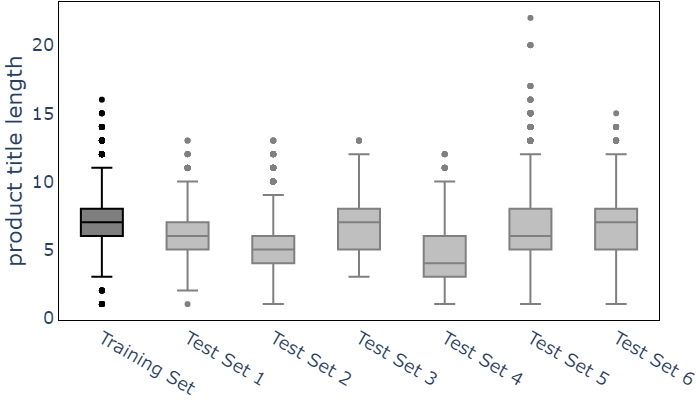}
    ~\caption{The distribution of product title lengths}
    \label{fig:boxplot}
\end{figure}

Figure~\ref{fig:boxplot} shows the distribution of the products' textual information, which is used as the independent feature in the classification models. 
The textual data follow a similar length pattern and are mostly short. 
The average number of words used in each product title is 6.2. 
It is worthwhile to note that having fewer vocabularies may degrade the model performance and make the learning process more difficult.

\begin{figure}[!ht]
  \centering
    \includegraphics[width=\linewidth]{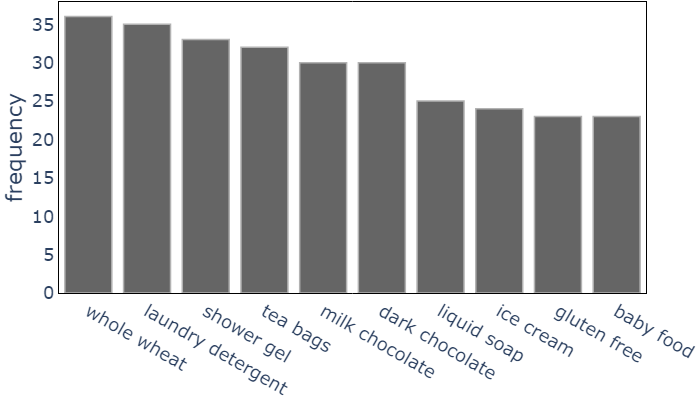}
    ~\caption{The most frequent bigrams in the product titles}
    \label{fig:bigram}
\end{figure}

Figure~\ref{fig:bigram} demonstrates the most frequent bigrams of product titles together with their frequencies in the training set. 
We observe that items such as bread, detergents, teas, and chocolates are the most frequent ones. 
Moreover, our exploratory data analysis reveals that 91 bigrams are repeated at least 10 times in the training set. 
These observations indicate a pattern in the titles that might facilitate the learning process for the classification models.

\subsection{Baseline Text Classification Models}
In our analysis, we consider various traditional text classification models such as XGBoost, SVM, and LSTMs, which we briefly summarize below.

\paragraph{XGBoost with Weighted Word Embedding} 
XGBoost, a scalable tree boosting method~\cite{Chen2016}, creates a group of weak trees by adding instances with the highest contribution to the model's learning process. Textual information cannot be directly used by the XGBoost, and they need to be converted to numeric values. TF-IDF is frequently used to this end. However, frequency-based approaches overlook the semantics and syntax of the vocabularies. Therefore, as suggested by~\citet{Stein2019}, we use word embeddings as the numeric representation, and apply a weighted average of the vocabularies given a product title, where weights are the TF-IDF of each word~\cite{jahanshahi2021}. 
While using TF-IDF as weights, we aim to give higher weights to the more important vocabularies. 
In our preliminary analysis, we compared this representation with TF-IDF only version as well as the simple word embedding average, and found it to be more efficient with a better overall performance. 
Moreover, we experiment with different word embeddings to identify the best-performing approach to convert the textual information given their context.

\paragraph{SVM with Weighted Word Embedding}
Support Vector Machine (SVM) is a widely used text classification method in different domains~\cite{Wibowo2018, Wang2017,goudjil2018}. 
Support vectors (i.e., data points that are closer to the hyperplane) are selected in a way that the classifier's margin is maximized. 
This model is able to independently learn feature space dimensions and can be used without feature selection.
Similar to XGBoost, SVM needs a numeric representation of the vocabularies. 
In our experiments, we employ the same approach, i.e., word embedding weighted average using TF-IDF, for the sake of consistency.

\begin{figure}[!ht]
  \centering
  \subfloat[Bidirectional LSTM]{
    \includegraphics[width=.32\linewidth]{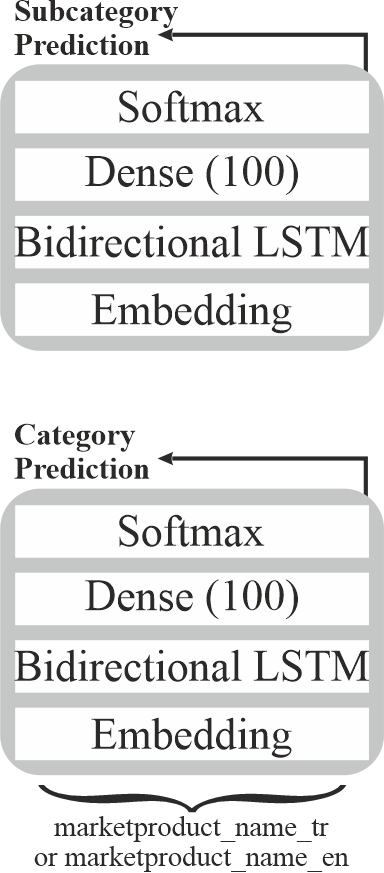}\label{fig:BiLSTM}
  } \enspace\enspace\enspace
  \subfloat[Bilingual Bidrectional LSTM]{
    \includegraphics[width=.62\linewidth]{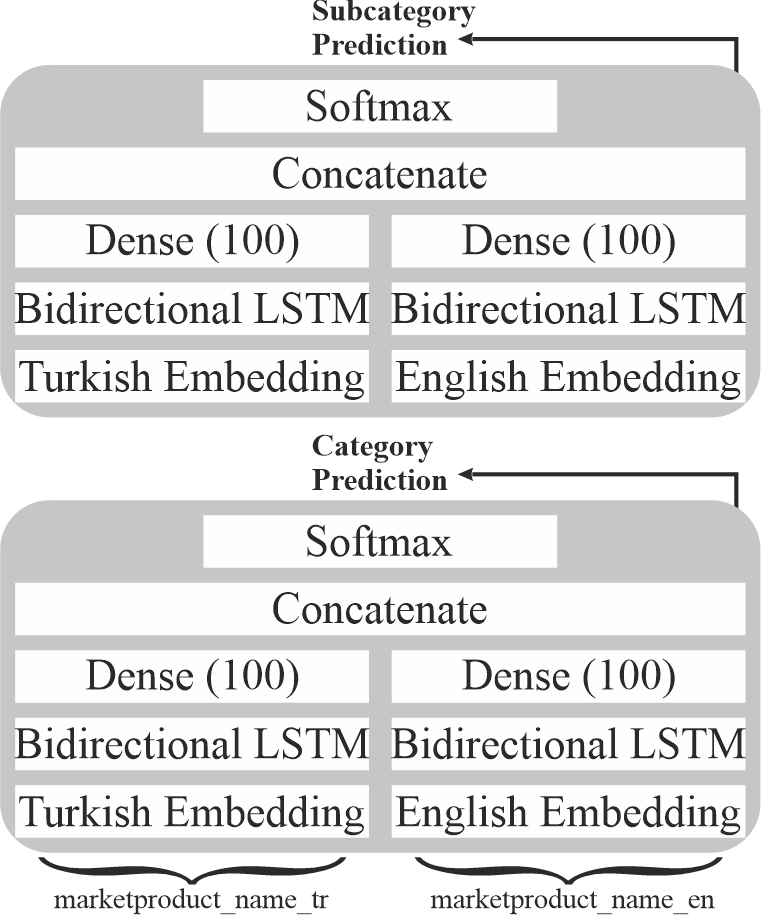} \label{fig:bilingual}
  }
    ~\caption{BiLSTM models used for multi-class classification}
    \label{fig:LSTM}
\end{figure}

\paragraph{Bi-directional LSTM}
Long short-term memory networks (LSTM), which is a particular type of recurrent neural network, can capture both long- and short-term effects of the textual information using the input, output, and forget gates~\cite{Liu2019}. 
LSTM's ability of when to learn new or relevant information and when to forget old or irrelevant information makes it a suitable tool in textual classification tasks. 
Since the title of a product can be lengthy and may include less relevant information, forget gates can filter out this kind of information. 
In this study, we use Bi-directional LSTM (BiLSTM) units, which learn the textual information from both directions, and then combine it into a single expression using the convolutional neural networks~\cite{Chenbin2018}. 
In our analysis, we employ two different LSTM architectures: one for predicting the labels using Turkish titles and another bilingual parallel LSTM which is fed by both Turkish and English titles (see Figure~\ref{fig:LSTM}). 
We use two independent networks for category and subcategory prediction. 
Their inputs are market product names in English and Turkish. 
Accordingly, the proper word embedding language is used.
However, their output dimensions differ and are equal to the number of categories and subcategories, respectively. 
Therefore, they are provided the product name and expected to return their associated category and subcategory. 
In this approach, we have two separate models for English and Turkish titles.

Figure~\ref{fig:bilingual} demonstrates our proposed parallel LSTM network used for bilingual prediction. 
The network has two separate word embeddings per language, and each LSTM network is updated separately.
Finally, by concatenating the result of Turkish and English LSTM networks, the category or subcategory of a product is predicted. 
Similarly, we have two networks for category and subcategory prediction. 
Note that we use this bilingual model only if the platform (from which the dataset is extracted) supports both languages. 

\subsection{Pretrained Language Models} 
For all the large pretrained language models (e.g., BERT, XLM, XLM-RoBERTa), we fine-tune the models by attaching two fully-connected layers to the output of the base models for the category and the subcategory. 
The contextual representation of the model input (e.g., product title) is obtained by concatenating the hidden layers of the base model. 
This vector is then provided to the fully-connected layers with Softmax activations. 
The fully-connected layers make the predictions for the category and the subcategory.
To further improve the performance of the models, we apply Dynamic Masked Softmax~\cite{tencent2020mwpd} on the activation of the fully-connected layer for the subcategory.
We consider three large multi-lingual pretrained language models, namely, BERT, XLM, and XLM-RoBERTa.

\paragraph{BERT} is a popular language representation model designed to pre-train deep bidirectional representations by jointly conditioning on both left and right context in all layers. 
The proposed architecture can be fine-tuned by only adding an output layer to generate state-of-art NLP models for a variety of tasks. 
For our experiments, we use the multi-lingual version of this architecture pretrained on 102 languages with the largest Wikipedia using a masked language modeling (MLM) objective~\cite{devlin2018bert}.

\paragraph{XLM}
\citet{lample2019crosslingual} propose an unsupervised method for learning cross-lingual representations using cross-lingual language modeling. 
The pretrained language model (XLM) that they design uses the Translation Language modeling (TLM) objective on top of BERT's Masked Language modeling (MLM) objective.
It consists of concatenating a sentence in two different languages with random masking. 
To predict one of the masked tokens, the model can use both the surrounding context in the first language as well as context given by the second language. 
The TLM objective allows the XLM model to learn better cross-lingual word embeddings. 
For our experiments, we use the multi-lingual version of this architecture pretrained on 102 languages with the largest Wikipedia using a masked language modeling (MLM) and Translation Language modeling (TLM) objectives~\cite{devlin2018bert}.

\paragraph{XLM-RoBERTa}
\citet{liu2019roberta} propose a variety of enhancements over the original BERT architecture and achieve better results on various NLP benchmark datasets. 
Their primary modifications for the BERT model include using additional datasets, changing some initial hyperparameters, removing next-sentence pretraining objectives, and training with larger batch sizes. 
For our experiments, we use the multi-lingual version of this architecture pretrained on 2.5TB of CommonCrawl data in 100 languages using a masked language modeling (MLM) objective~\cite{wolf2019huggingface}.

\subsection{Dynamic Masking}
\begin{figure}[!ht]
  \centering
    \includegraphics[width=.4\textwidth]{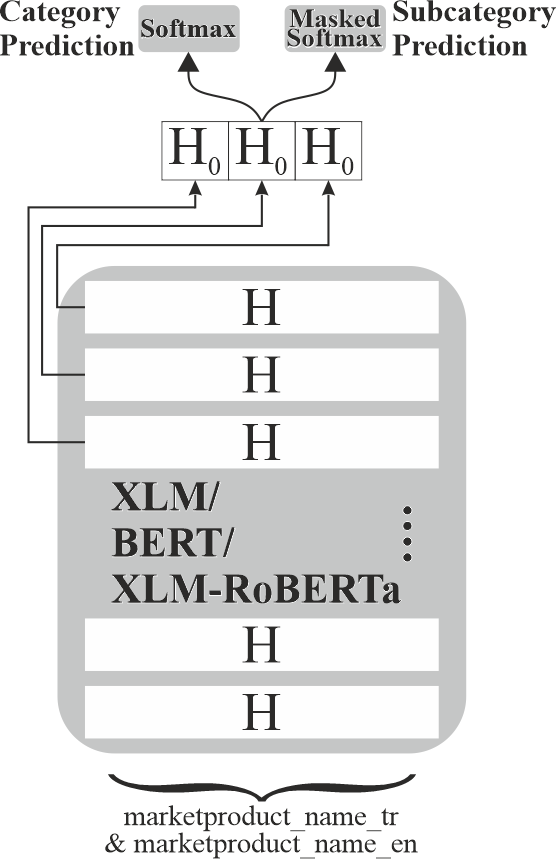}
    ~\caption{Masked XLM, XLM-RoBERTa, and Bert~\cite{tencent2020mwpd}}
    \label{fig:masked}
\end{figure}

The architecture we employed for multi-level classification using pre-trained language models is illustrated in Figure~\ref{fig:masked}.
In the standard approach, the hidden states of the base model ($H$) are concatenated to obtain an encoded representation of the inputs. 
Then, two feed-forward neural layers with Softmax activations are applied to compute the probabilities of categories for each level. 
However, the standard Softmax layer performs suboptimally, ignoring the dependencies between categories and subcategories.

An alternative method proposed by \citet{tencent2020mwpd} uses Dynamic Masked Softmax to dynamically filter out the child categories which are unrelated to the current parent category. 
For instance, if the category $c_i$ is predicted by the model, the model should only recommend subcategories that fall in the same category. 
Hence, infeasible subcategories should be filtered out using a binary mask. 
In this method, we define a binary mask matrix $M \in \{0,1\}^{C \times S}$, where $C$ is the number of categories and $S$ is the number of subcategories. 
We then compute the Dynamic Masked Softmax instead of regular Softmax for computing the subcategory predictions as follows:
\begin{equation}
    P(y_s|c,\theta) = \frac{exp(O_s) M_{c,s}+exp(-8)}{\sum_{s'=1}^S exp(O_{s'}) M_{c,s'}  + exp(-8) }
\end{equation}
where $c$ and $s$ are category and subcategory labels, $\theta$ is the model parameters, and $y_s$ is the predicted probabilities of the subcategories.
This design can also be extended to more than two levels if needed. 
In our numerical analysis, we experiment with different configurations to measure the impact of Dynamic Masked Softmax using training three language models.

\subsection{Experimental Setup}
\begin{figure}[!ht]
    \centering
    \includegraphics[width=.9\linewidth]{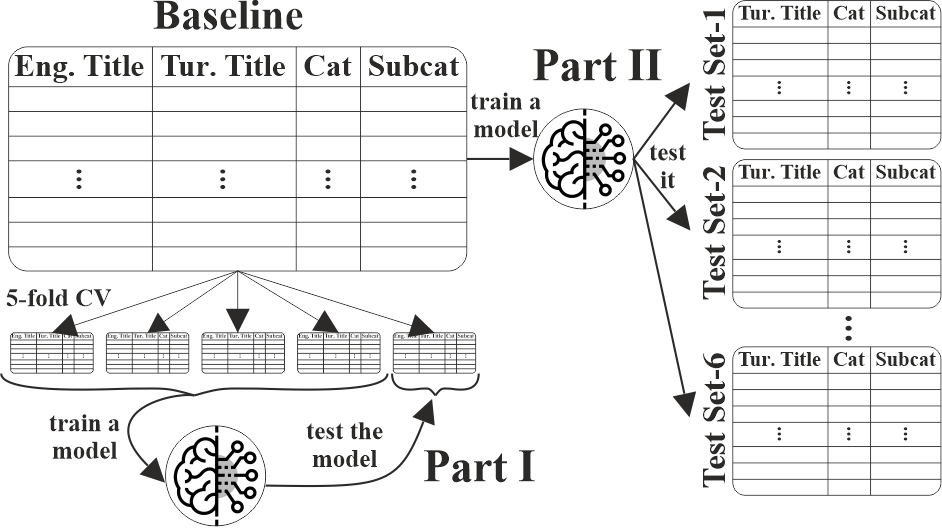}
    \caption{Experimental setup}
    \label{fig:process}
\end{figure}

Our experimental setup is illustrated in Figure~\ref{fig:process}, which consists of two parts. 
In the first part, we apply five-fold cross-validation to the training set described in Section~\ref{tab:datasets}. 
At this stage, we perform experiments to measure the category and subcategory prediction accuracy for different models and word embeddings. 
Additionally, we investigate the advantage of using bilingual product titles and, finally, the benefit of applying dynamic masking of subcategories for the pretrained language models.
In the second part, we take the best models trained on the training data and evaluate their performance on the test sets to measure the generalizability of the trained models.

\paragraph{Evaluation Metrics}
We use two generic metrics to assess the model performance: accuracy and weighted-average macro F1-score (WAF1).
Accuracy is an easy-to-interpret metric that shows how often the model is correct. 
We compute the accuracy for both category and subcategory predictions. 
However, in comparing the models, we consider the weighted-average F1-scores calculated over all the categories for each classification level. 
We compute the F1-scores using the harmonic mean of precision and recall scores as follows:
\begin{equation*}
    \text{F1-score}=2 \times \frac{\text { precision } \times \text { recall }}{\text { precision }+\text { recall }}
\end{equation*}
F1-scores are calculated for each class independently, and a weighted average is taken to obtain WAF1 scores.
We rank the models using the same aggregated metric proposed by \citet{zhang2020mwpd2020}, i.e., by averaging the WAF1 for the category and the subcategory.

\paragraph{Parameter Settings}
For the implementation of the BiLSTM and pretrained language models, we use the Tensorflow and Transformers~\cite{wolf-etal-2020-transformers} libraries. 
We fine-tune the \textit{bert-base-multilingual-uncased}, \textit{xlm-mlm-100-1280}, and \textit{jplu/tf-xlm-roberta-base} versions of the pre-trained language models in Transformers library for BERT, XLM, and XLM-RoBERTa, respectively. 
For all the pre-trained language models, we use Adam optimizer with an initial learning rate of 3e-05, and a batch size of 16.
During the training of each model, early-stopping is applied to avoid over-fitting. 
Training is stopped when no performance improvement is observed on the validation set after 10 epochs. 
We then store the model weights corresponding to the best performance on the validation set. 

We use a grid search to fine-tune the parameters of the SVM, XGBoost, and BiLSTM as well. 
This process is done using a separate validation set as discussed in sections~\ref{sec:self-prediction} and \ref{sec:transfer-learning}. 
We use the scikit-learn library in Python for these implementations. 
Table~\ref{tab:parameters} lists the hyperparameters used for each model.

\renewcommand{\arraystretch}{1.2} 
\begin{table}[!ht]
    \caption{Default model parameters used in the experiments}
    \centering
    \resizebox{\linewidth}{!}{
    \begin{tabular}{ll}
    \toprule
    \textbf{Model} & \textbf{Hyperparameters} \\
     \midrule
    \textbf{SVM} & \begin{tabular}[l]{@{}l@{}}kernel = linear, degree of polynomial kernel= 3\\ $C = 1.0$\end{tabular} \\
    \textbf{XGboost} & \begin{tabular}[l]{@{}l@{}}number of trees = 100, max depth = 3, \\ learning rate = 0.1 \end{tabular} \\
    \textbf{LSTM} & \begin{tabular}[l]{@{}l@{}} Embedding layer, BiLSTM layer = 200, \\ Dense layer = (100, relu activeation), \\ Dense layer = (n of cat/subcat, softmax activation) \\ optimizer= adam, \#epochs=20 \end{tabular}\\
    \textbf{BERT,  XLM,  XLM-RoBERTa} & \begin{tabular}[l]{@{}l@{}} learning rate=3e-5, batch size=16 \\ early stopping patience=10,  \#epochs=100 \end{tabular}\\
     \bottomrule
    \end{tabular}
    }
    \label{tab:parameters}
\end{table}

\section{Numerical Results}\label{sec:experiments}
We focus on two particular sets of experiments: finding the best model using cross-validation on the baseline dataset and measuring generalizability on multiple datasets mined from various other online retail stores.
We obtained our datasets from Getir, an online food and grocery delivery company that originated in Turkey and recently expanded its operations to the United Kingdom and the Netherlands.
Since Getir primarily operates in Turkey, the product titles they collected usually have Turkish titles.
As such, classification models can significantly benefit from multilingual word embeddings.
In our numerical study, we first explore the performance of the baseline models together with recent deep learning approaches and choose top classifiers for the next step. 
Then, we utilize those models to predict the categories and subcategories of the other online grocery retailers. 
Finally, we provide detailed insights on the model performance, i.e., where the model fails in their prediction and how different word embeddings affect their performance.

\subsection{Performance comparison}\label{sec:self-prediction}
The baseline dataset provides full information related to 3,228 unique products that are sold online. 
For this experiment, we use the products' Turkish and English titles, categories, and subcategories. 
We take advantage of the availability of bilingual product titles and investigate different word embedding approaches. 
We also utilized a complete list of multi-class classification algorithms, including the more traditional machine learning algorithm, e.g., SVM and XGBoost, together with more recent deep learning algorithms, e.g., LSTM and BERT. 

Using the stratified cross-validation, we split the dataset into 5 folds, where one fold is used as the test set and the rest as the training set. 
This process is repeated five times for all folds. 
Using a 90-10 division, we create the validation set out of the training set and optimize the model parameters accordingly. 
This approach enables finding the optimal parameters and avoid overfitting. 
The self-prediction phase leads to a list of possible models to be applied in the cross-platform multilevel classification task.

Table~\ref{tab:model_performance} shows the performance of each model for 5-fold cross-validation given the applied word embedding. 
As the dataset includes both English and Turkish titles, we consider a combination of Turkish, English, and Multilingual word embeddings. 
Examining the accuracy and F1-score of the models, we conclude that Turkish Glove Embedding, in comparison to English Glove, Turkish FastText, and Turkish Word2Vec embeddings, provides a better numeric representation for the vocabularies used in our dataset. 
We note that bidirectional LSTM with Turkish Glove embedding has the highest accuracy compared to other traditional baseline models. 
We also explore the feasibility of a bilingual LSTM model to predict the product categories and subcategories. 
We observe that using a parallel BiLSTM, which is input by both Turkish and English titles, further enhances the accuracy and F1-score of the classifier.
For the pretrained language models trained using Turkish titles, we find that the BERT model performs best, followed by the XLM model and the XLM-RoBERTa model. 
This finding is consistent with the recent literature in product category classification that uses BERT architecture over other alternatives~\cite{zhang2020mwpd2020, tencent2020mwpd}. 
We also note that dynamic masking of subcategories increases the model performance of all the pretrained language models. 
However, the use of bilingual titles only increases the accuracy of the BERT architecture. 
These results are discussed further in Section~\ref{results-masking} and Section ~\ref{results-bilingual}. 
Overall, the results show that BERT architecture with bilingual titles and dynamic masking performs best, followed by the Bi-LSTM model with bilingual titles.

\setlength{\tabcolsep}{4pt} 
\renewcommand{\arraystretch}{1.2} 
\begin{table*}[!ht]
    \caption{Comparison of different models and word embedding approaches} \label{tab:model_performance}
    \resizebox{\linewidth}{!}{
    \begin{tabular}{llccccr}
        \toprule
        \multirow{2}{*}{\textbf{Model}} & \multirow{2}{*}{\textbf{Word Embedding}} & \multicolumn{2}{c}{\textbf{Accuracy (\%)}} & \multicolumn{3}{c}{\textbf{Weighted Avg. F1-Score (\%)}} \\
         &  & \textbf{Cat} & \textbf{Sub} & \textbf{Cat} & \textbf{Sub} & \textbf{Avg.} \\
         \midrule
        SVM & \multirow{3}{*}{Turkish Word2Vec} & 85.0 ± 0.8  & 74.2 ± 1.5 & 84.7 ± 0.9 & 72.4 ± 1.7 & 78.6 \\
        XGBoost &  & 79.0 ± 1.1 & 70.6 ± 0.8 & 78.4 ± 1.2 & 69.3 ± 1.0 & 73.9 \\
        Bi-LTSM &  & 94.5 ± 0.9 & 87.1 ± 1.4 & 94.5 ± 0.9 & 86.7 ± 1.4 & 90.6 \\
        \midrule
        SVM & \multirow{3}{*}{Turkish FastText} & 78.4 ± 1.0 & 62.8 ± 2.4 & 78.7 ± 0.9 & 57.7 ± 3.3 & 68.2 \\
        XGBoost &  & 85.3 ± 1.4 & 77.2 ± 1.8 & 84.7 ± 1.4 & 75.8 ± 1.8 & 80.3 \\
        Bi-LTSM &  & 95.3 ± 0.8 & 89.0 ± 1.0& 95.3 ± 0.8& 88.7 ± 0.9 & 92.0 \\
        \midrule
        SVM & \multirow{3}{*}{Turkish Glove} & 93.6 ± 1.2& 88.8 ± 2.2 &93.5 ± 1.2 &88.0 ± 2.4 & 90.8 \\
        XGBoost &  & 90.5 ± 1.1& 82.7 ± 1.4& 90.2 ± 1.2& 81.6 ± 1.6 & 85.9 \\
        Bi-LTSM &  & 96.4 ± 0.8 & 91.1 ± 1.2& 96.4 ± 0.9& 90.7 ± 1.2 & 93.6 \\
        \midrule
        SVM & \multirow{3}{*}{English Glove} & 80.1 ± 1.7 &79.2 ± 2.9 &79.7 ± 1.8 &77.7 ± 3.3 & 78.7 \\
        XGBoost &  &  78.4 ± 1.5& 71.9 ± 2.0& 77.5 ± 1.4 & 70.3 ± 2.0 & 73.9 \\
        Bi-LTSM &  & 92.9 ± 1.3& 86.5 ± 1.4& 92.8 ± 1.4& 86.3 ± 1.5 & 89.6 \\
        \midrule
        Bi-LTSM & Bilingual (Tur. \& Eng. Glove) & \textbf{96.6 ± 0.4} & 91.2 ± 0.6 & \textbf{96.6 ± 0.4} & 90.9 ± 0.6 & 93.8 \\
        \midrule
        BERT & Multilingual, Turkish titles & 95.9 ± 1.2 & 91.5 ± 1.4 & 95.8 ± 1.2 & 91.2 ± 1.5 & 93.5 \\
        XLM & Multilingual, Turkish titles & 96.0 ± 0.8 & 89.0 ± 1.7 & 95.9 ± 0.9 & 88.5 ± 1.9 & 92.2 \\
        XLM-RoBERTa & Multilingual, Turkish titles & 95.2 ± 1.2 & 88.3 ± 1.0 & 95.2 ± 1.1 & 87.9 ± 0.9 & 91.6 \\ 
        \midrule
        BERT$^\mathcal{M}$ & Multilingual, Turkish titles & 95.2 ± 0.9 & 93.1 ± 0.6 & 95.2 ± 0.8 & 92.8 ± 0.6 & 94.0 \\
        XLM$^\mathcal{M}$ & Multilingual, Turkish titles & 96.0 ± 0.8 & 92.0 ± 1.8 & 95.9 ± 0.8 & 91.3 ± 2.1 & 93.6 \\
        XLM-RoBERTa$^\mathcal{M}$ & Multilingual, Turkish titles & 95.4 ± 1.0 & 92.7 ± 2.4 & 95.3 ± 1.0 & 92.2 ± 2.6 & 93.8\\ 
        \midrule
        BERT$^\mathcal{M}$ & Multilingual, Bilingual titles & 95.4 ± 0.7 & \textbf{93.9 ± 0.9} & 95.3 ± 0.8 & \textbf{93.7 ± 1.0} & \textbf{94.5} \\
        XLM$^\mathcal{M}$ & Multilingual, Bilingual titles & 95.0 ± 1.0 & 90.9 ± 0.8 & 94.8 ± 1.0 & 90.3 ± 0.9 & 92.5 \\
        XLM-RoBERTa$^\mathcal{M}$ & Multilingual, Bilingual titles & 94.9 ± 1.6 & 90.9 ± 1.6 & 94.7 ± 1.6 & 90.4 ± 1.7 & 92.6 \\
        \bottomrule
    \end{tabular}
    }
    \begin{tablenotes}
        \footnotesize
        \item \hspace{2.2cm} $^\mathcal{M}$: Masked version for subcategory prediction.
    \end{tablenotes} 
\end{table*}

\subsection{Impact of Dynamic Masking}\label{results-masking}
For the pretrained language models, we investigate the benefits of dynamic masking on multi-level product category classification. 
We train three pretrained language models with and without masking and measure category and subcategory prediction performance. 
The results are provided in Table~\ref{tab:model_performance}, where the models with dynamic masking have the superscript $\mathcal{M}$ (e.g., BERT$^\mathcal{M}$, XLM$^\mathcal{M}$, and {XLM-RoBERTa$^\mathcal{M}$}).

We also train the models on Turkish titles with and without the masking.
Table~\ref{tab:model_performance} shows that masked models always lead to better overall performance. 
This observation is intuitive since masking simplifies the classification task by reducing the number of subcategory classes to those under the predicted category. 
Thus, we observe that the category accuracy remains similar while the subcategory accuracy increases when the mask is employed.

\subsection{Impact of Incorporating Bilingual Titles}\label{results-bilingual}
When the dataset contains product titles in multiple languages, it is possible to leverage this information for better prediction performance. 
Thus, we train BiLSTM, BERT$^\mathcal{M}$, XLM$^\mathcal{M}$, and XLM-RoBERTa$^\mathcal{M}$ with multiple titles as well. 
For the BiLSTM model, we use the Glove embeddings, which perform better than Word2Vec and FastText embeddings. 
The BiLSTM model that uses both English and Turkish embeddings performs significantly better than the BiLSTM models that only use Turkish or English embeddings (see Table~\ref{tab:model_performance}).

For bilingual pretrained language models, the Turkish and English product titles are concatenated and provided to the model. 
This concatenation is possible since the pretrained models are multilingual and support 100 languages, including Turkish and English. 
Interestingly, using bilingual titles with pretrained models does not lead to consistent performance improvements. 
While we observe a clear accuracy improvement for both category and subcategory for the BERT$^\mathcal{M}$ model, XLM$^\mathcal{M}$ and XLM-RoBERTa$^\mathcal{M}$ perform worse when both product titles are used. 
The reason why we do not observe the same improvements in the XLM$^\mathcal{M}$ and XLM-RoBERTa$^\mathcal{M}$ models might be due to using the Translation Language modeling (TLM) objective in pre-training. 
This training objective allows the model to learn better cross-lingual word representations. 
Thus, feeding bilingual titles might not lead to an improvement. 

\subsection{Cross-platform Multilevel Classification}\label{sec:transfer-learning}
As a market analysis, we crawl data related to six online retailers. 
The description of each dataset after data cleaning is reported in Table~\ref{tab:datasets}.
Not all online grocery shops have an English version. 
Therefore, we use only Turkish titles to be consistent for all the platforms. 
Moreover, in the mined datasets, some categories and subcategories do not exist in the baseline dataset. 
Accordingly, we only consider the products whose categories and subcategories exist on our baseline dataset. 
We use 80 percent of our dataset as the training set and the rest as the validation set to identify ideal model parameters. 
We then examine the feasibility of cross-platform multilevel product category classification.

BiLSTM, BERT$^\mathcal{M}$, XLM$^\mathcal{M}$, and {XLM-RoBERTa$^\mathcal{M}$} are selected as the top models in our experiments for within-platform classification. 
Table~\ref{tab:transfer_learning} summarizes the performance of the models for each dataset where the best classifiers are bolded. 
{XLM-RoBERTa$^\mathcal{M}$} is not the best model in any of the cases. 
Surprisingly, Bidirectional LSTM is on a par with the masked version of the BERT and XLM. 
For test set - 1, masked models are significantly better in category prediction, whereas BiLSTM surpasses in subcategory prediction. 
Masked versions are the best classifiers for test set - 2, while for the third one, BiLSTM outperforms in terms of category prediction. 
XLM$^\mathcal{M}$ is the best predictor for test set - 4, and BiLSTM has the highest performance for test set - 6. 
In predicting the product categories for the fifth test set, BiLSTM better predicts categories, and BERT$^\mathcal{M}$ is substantially better in subcategory prediction. 
Overall, we note that there is no single remedy for a product category or subcategory prediction. 
Therefore, we recommend applying an ensemble of models for future steps. 
Also, Stochastic Weight Averaging (SWA) can be leveraged to improve the performance of the models in imbalanced product labels.
Based on the cross-platform classification results, we remark that the models trained on our original dataset are feasible to be generalized to the datasets obtained from other online retailers. 
We note that the product titles in each test sets are different. 
As this difference becomes more substantial when compared to the training set, the classifier performance decreases further.



\begin{table*}[!ht]
    \caption{F1-score reported for the generalizability to competitors' products} \label{tab:transfer_learning}
    \resizebox{\linewidth}{!}{
    \begin{tabular}{lrrrrrrrrrrrrrrrrrr}
        \toprule
        \multirow{2}{*}{\textbf{Models}} & \multicolumn{3}{c}{\textbf{Test Set - 1}} & \multicolumn{3}{c}{\textbf{Test Set - 2}}& \multicolumn{3}{c}{\textbf{Test Set - 3}} & \multicolumn{3}{c}{\textbf{Test Set - 4}} & \multicolumn{3}{c}{\textbf{Test Set - 5}} & \multicolumn{3}{c}{\textbf{Test Set - 6}} \\
         & \textbf{Cat} & \textbf{Subcat} & \textbf{Avg.} & \textbf{Cat} & \textbf{Subcat} & \textbf{Avg.} & \textbf{Cat} & \textbf{Subcat} & \textbf{Avg.} & \textbf{Cat} & \textbf{Subcat} & \textbf{Avg.} & \textbf{Cat} & \textbf{Subcat} & \textbf{Avg.} & \textbf{Cat} & \textbf{Subcat} & \textbf{Avg.} \\
         \midrule
        \textbf{BiLSTM} & 89.2 & \textbf{84.3} & 86.7 & 82.1 & 71.8  & 77.0 & \textbf{97.0} & 93.8 & 95.4 & 74.3 & 62.6 & 68.5 & \textbf{90.2} & 87.2 & 88.7 & \textbf{85.0} & \textbf{82.1} & \textbf{83.6} \\
        \textbf{BERT$^\mathcal{M}$} & \textbf{95.3} & 78.3 & \textbf{86.8} & 82.0 & 76.5 & 79.3 & 96.3 & \textbf{95.1} & \textbf{95.7} & 75.6 & 68.8 & 72.2 & 86.9 & \textbf{90.7} & \textbf{88.8} & 81.2 & 78.4 & 79.8 \\
        \textbf{XLM$^\mathcal{M}$} & 91.3 & 75.3 & 83.3 & \textbf{82.9} & \textbf{78.7}  & \textbf{80.8} & 95.7 & 94.5 & 95.1 & \textbf{77.9} & \textbf{72.3} & \textbf{75.1} & 87.7 & 78.1 & 82.9 & 81.8 & 77.7 & 79.8 \\
        \textbf{XLM-RoBERTa$^\mathcal{M}$} & 90.1 & 69.5 & 79.8 & 80.6 & 74.3  & 77.4 & 95.4 & 90.5 & 92.9 & 74.3 & 67.5 & 70.9 & 86.9 & 67.4 & 77.1 & 78.6 & 73.4 & 76.0 \\
        \bottomrule
    \end{tabular}
    }
    \begin{tablenotes}
        \footnotesize
        \item $^\mathcal{M}$: Masked version for subcategory prediction.
    \end{tablenotes}    
\end{table*}

\subsection{Discussion on Model Predictions}\label{sec:discussions}

To investigate the products in other (test) datasets for which the models fail to predict the category or subcategory, we visually compare the predicted values with the ground truth. 
Our observations of the misclassifications are as follows.
\begin{itemize}
    \item If a product does exist in the test set but not in the training set or has different wording than the training set, a misclassification may occur. 
    
    \item Some brand names have a general meaning that affects the model's prediction. 
    For instance, the manufacturer ``doğuş'', meaning ``nativity'' in Turkish, produces beverage products, while its meaning conveys a different understanding for the prediction models. 
    
    \item Product categories can be subjective. For instance, a company categorizes a product as a dairy product, whereas another one sets its category as beverages. This issue cannot be addressed in the cleaning phase as we deal with an extensive list of product names and categories in this work. On the other hand, a manual check may still have the subjectivity mentioned. Therefore, we rely on the category naming as is. 
    
    \item Some product titles convey a meaning that can be taken differently by a machine learning algorithm. For instance, a book name that is about cooking might be categorized as food. 
\end{itemize}

Table~\ref{tab:Misprediction} lists some products where our proposed model fails to predict the exact category or subcategory. 
``Report to El Greco'' is the name of a book and is categorized in Newspapers and Magazines; however, the model suggests paper products as its subcategory. 
Using the word ``report'' justifies this suggestion. 
Doğuş company is famous for its diverse tea products in Turkey. 
Therefore, the model associates it with the tea subcategory, even when the word ``sugar'' exists in the product name. 
Granola includes almond and cashew that also commonly exist in snacks. 
Even though the model misclassifies the product, it might still be considered as a logical prediction. 
Moreover, high protein vanilla milk clearly belongs to the milk subcategory as suggested by the model. 
However, it was originally categorized as ``Fitness and Form''. 
We investigate the rationale behind such a category selection and find that this product includes zero sugar, has high protein, and is lactose-free. 
Therefore, the retailer decided to categorize it as a dairy product under the ``fit \& form'' category. 
Lastly, the model categorizes ``Menemen mixture'', prepared ingredient for Turkish omelet, as ``Canned and pickled'', whereas it was originally categorized as ``Spices''. 
The rationale behind the new proposed subcategory is that this mixture also is sold in a jar. 
The model suggests an acceptable alternative for the current version without being aware of this fact. 
This evidence corroborates the subjectivity in naming conventions.
In this regard, including more details on the product content than what the pure title suggests might be considered as a viable strategy to improve the prediction performance. 
Overall, we note that although there are some cases where the model is unable to give the exact category or subcategory name, the predicted values are justifiable given the input provided to the model. 

\begin{table*}[!ht]
    \caption{Sample examples of mispredicted product categories and subcategories. Original titles and (sub)category names are translated from Turkish to English.} \label{tab:Misprediction}
    \resizebox{\linewidth}{!}{
        \begin{tabular}{lllll}
        \toprule
        \textbf{Product Title} & \textbf{Category} & \textbf{Category Prediction} & \textbf{Subcategory} & \textbf{Subcategory Prediction} \\
        \midrule
        \textbf{Report to El Greco} & Home \& Living & Home Care & Newspaper \& Magazine & Paper Products \\
        \textbf{\textit{Doğuş} sugar 1 kilogram} & Basic Food & Water \& Drink & Sugar & Tea \\
        \textbf{\textit{Granolife} almond and cashew granola 75 grams} & Fit \& Form & Snack & Granola & Cake \& Biscuit \\
        \textbf{\textit{Danone} pro+ high protein vanilla milk 330 milliliter} & Fit \& Form & Milk \& Breakfast & Dairy Products & Milk \\
        \textbf{\textit{Tat} Menemen (spicy Turkish omelette) mixture} & Basic Food & Food & Spices & Canned \& Pickled \\
        \bottomrule
        \end{tabular}
    }
\end{table*}

\section{Threats to Validity}\label{sec:threats}
In this study, we evaluated a comprehensive list of text classification techniques to address the multilevel product category classification task. 
We ensure to cover both well-established and novel approaches in this area. 
However, NLP is a fast-developing field,
and we aim to closely follow the trends and apply other methods for our prediction task in the future.
Moreover, an in-depth grid search for parameter tuning of the current models might prove fruitful.
In terms of construct validity, we used repeated stratified 5-fold cross-validation to alleviate the issue of random heterogeneity of subjects. 

Regarding the external sources, we mined datasets from six different online retail platforms in Turkey with distinct characteristics to ensure the generalizability of our findings. 
Note that we cover the most successful online retail companies in the field. 
Nonetheless, replication of our study for other languages and other countries might yield fruitful insights. 
Moreover, we consider only the title of products for this task. 
This information can be expanded by adding product descriptions, specifications, and prices. 
Incorporating additional information is set as a future step to enhance predictive performance. 


The datasets for other companies were extracted once a month from October 2019 to January 2021. We consider a unique list of products that are available during this period. However, some products might be out-of-stock while crawling the websites or not feasible to non-user viewers. Therefore, it is important to note that our analysis applies to the products that were open to public access during those days.

\section{Conclusion}\label{sec:conclusion}
When exploring marketing strategies, companies do not typically have access to full information about the products available in the marketplace. 
As such, they need to predict the missing information and match them with their category definitions to have a better sense of the market. 
In addition, the companies may aim to identify incorrectly classified products based on the products existing in their database to understand the recent market trends. 
In this paper, we studied the text classification strategies to automate the prediction of product categories and subcategories using the available information such as product titles. 
We analyzed the mined datasets related to top online grocery platforms in Turkey and utilized different machine learning algorithms to address the problem. 
We also designed a bilingual deep learning architecture that uses both English and Turkish product titles. 
After comparing the result of the models, we investigated the cases where the models fail to predict the expected categories, which can be particularly useful to pinpoint the cases where the current ground truth labels (i.e., categories) might be controversial. 

We plan to extend this research by adding additional information on the products, e.g., description, price, and ingredients, to enhance the predictive performance. 
A relevant venue for future research would be designing strategies to achieve better performance for certain categories. For instance, the trained models had low accuracy in the ``Newspaper \& Magazine" subcategory. Pre-training on a dataset about books or training an additional book/non-book classifier can increase the performance for this category without sacrificing performance on other categories.
Finally, the similarity of some categories presents a problem both to the models and to the practitioners. A refined strategy can be developed to quickly determine categories that are likely to contain very similar products. Based on this information, companies can more effectively categorize the products.

\section*{Acknowledgements}
The authors would like to thank the Getir company for supporting this study and providing data and feedback throughout.

\bibliographystyle{plainnat} 
\bibliography{Ref}
\end{document}